\documentclass[12pt]{article}
\usepackage[english,german,french,polish]{babel}
\usepackage[T1]{fontenc}
\usepackage{amsfonts}

\begin{document}

\selectlanguage{english}

\baselineskip 0.75cm
\topmargin -0.6in
\oddsidemargin -0.1in

\let\ni=\noindent

\renewcommand{\thefootnote}{\fnsymbol{footnote}}

\pagestyle {plain}

\setcounter{page}{1}

\pagestyle{empty}

~~~

\begin{flushright}
IFT--04/7
\end{flushright}

{\large\centerline{\bf Two light sterile neutrinos that mix maximally with 
each other}}

{\large\centerline{\bf and moderately with three active 
neutrinos{\footnote{Work supported in part by the Polish State Committee 
for Scientific Research (KBN), grant 2 P03B 129 24 (2003--2004).}}}}

\vspace{0.4cm}

{\centerline {\sc Wojciech Kr\'{o}likowski}}

\vspace{0.3cm}

{\centerline {\it Institute of Theoretical Physics, Warsaw University}}

{\centerline {\it Ho\.{z}a 69,~~PL--00--681 Warszawa, ~Poland}}

\vspace{0.6cm}

{\centerline{\bf Abstract}}

\vspace{0.2cm}

Since the 3+1 neutrino models with one light sterile neutrino turn out to be 
not very effective, at least two light sterile neutrinos may be needed to 
reconcile the solar and atmospheric neutrino experiments with the LSND result,
{\it if}  this is confirmed by the ongoing MiniBooNE experiment (and when the 
CPT invariance is assumed to hold for neutrino oscillations). We present an 
attractive 3+2 neutrino model, where two light sterile neutrinos mix maximally 
with each other, in analogy to the observed maximal mixing of muon and tauon 
active neutrinos. But, while the mixing of $\nu_e$ and 
$(\nu_\mu-\nu_\tau)/\sqrt2 $ is observed as large (though not maximal), the 
mixing of  $\nu_e$ with the corresponding combination of two light sterile 
neutrinos is expected to be only moderate because of the reported smallness 
of LSND oscillation amplitude. The presented model turns out, however, not 
to be {\it more} effective in explaining the hypothetic LSND result than the 
simplest 3+1 neutrino model. On the other hand, in the considered 3+2 model, 
the deviations from conventional oscillations of three active neutrinos appear 
to be {\it minimal} within a larger class of 3+2 models.

\vspace{0.2cm}

\ni PACS numbers: 12.15.Ff , 14.60.Pq , 12.15.Hh .

\vspace{0.6cm}

\ni February 2004 

\vfill\eject

~~~
\pagestyle {plain}

\setcounter{page}{1}

\vspace{0.2cm}

\ni {\bf 1. Introduction}. As is well known, the neutrino experiments with 
solar  $\nu_e$'s [1], atmospheric $\nu_\mu$'s [2] and long-baseline reactor 
$\bar{\nu}_e$'s [3] are very well described by oscillations of three active 
neutrinos $\nu_e \,,\, \nu_\mu \,,\, \nu_\tau $, where the mass-squared 
splittings of the related neutrino mass states $\nu_1, \nu_2 , \nu_3 $ are 
estimated to be $\Delta m^2_{\rm sol}\equiv\Delta m^2_{21}\sim7\times10^{-5}$
eV$^2$ and $\Delta m^2_{\rm atm}\equiv\Delta m^2_{32}\sim2\times 10^{-3}$ 
eV$^2$ [4]. The neutrino mixing matrix 
$U^{(3)}=\left(U^{(3)}_{\alpha i}\right)$ ~($\alpha= e$ $\mu$, $\tau$ and 
$i=1,2,3$), appearing in the unitary transformation

%rownanie 1
\begin{equation}
\nu_\alpha  = \sum_i U^{(3)}_{\alpha i}\, \nu_i \;,
\end{equation}

\ni is experimentally consistent with the global bilarge form

\vspace{-0.1cm}

%rownanie 2
\begin{equation}
U^{(3)} = \left( 
\begin{array}{ccc} c_{12} & s_{12} & 0 \\ 
- \frac{1}{\sqrt2} s_{12} & \frac{1}{\sqrt2} c_{12} & \frac{1}{\sqrt2}\\ 
\frac{1}{\sqrt2} s_{12} & -\frac{1}{\sqrt2} c_{12} & \frac{1}{\sqrt2}  
\end{array} \right) \;,
\end{equation}

\ni where $\theta_{12} \sim 33^\circ $ and $\theta_{23} = 45^\circ $, while 
$U^{(3)}_{e3} = s_{13} \exp(-i\delta)$ is neglected in accordance with the 
negative result of Chooz experiment with short-baseline reactor 
$\bar{\nu}_e$'s [5] (the experimental upper bound is estimated at 
$s_{13} < 0.03$). However, the signal of 
$\bar{\nu}_\mu \rightarrow \bar{\nu}_e $ oscillations reported by LSND 
experiment with short-baseline accelerator $\bar{\nu}_\mu$'s [6] requires a 
third neutrino mass-squared splitting, say, $\Delta m^2_{\rm LSND}\sim1$ eV$^2$
which cannot be justified by the use of only three neutrinos (unless the CPT 
invariance of neutrino oscillations is seriously violated, leading to 
considerable mass splittings of neutrinos and antineutrinos [7]; in the 
present note the CPT invariance is assumed to hold for neutrino oscillations). 
The LSND result will be tested soon in the ongoing MiniBooNE experiment [8]. 
If this test confirms the LSND result, we will need the light sterile 
neutrinos in addition to three active neutrinos to introduce extra mass 
splittings (and, at the same time, not to change significantly the solar, 
atmospheric and reactor neutrino oscillations).

While the 3+1 neutrino models with one light sterile neutrino are considered 
to be strongly disfavored by present data [9], the 3+2 neutrino schemes with 
two light sterile neutrinos may provide a much better description of current 
neutrino oscillations including the LSND effect (for a discussion on the 
compatibility of all short-baseline neutrino experiments in 3+1 and 3+2 models 
{\it cf.} Ref [10]).

The necessary existence in Nature of exactly two light sterile neutrinos was 
argued some years ago [11] on the ground of a new series of 
{\it generalized (K\"{a}hler-like) Dirac equations} which could describe 
three and only three generations of SM-active leptons and quarks, and two and 
only two generations of single SM-passive light neutrinos (light sterile 
neutrinos). The condition for it was an 
{\it intrinsic (Pauli-type) exclusion principle} that, when it was holding, 
cut off the series of the corresponding generalized Dirac fields to one triad 
of SM (15+1)-plets and one couple of SM singlets, respectively (all of spin 
1/2). The subjects of this exclusion principle were the sets of 
{\it additional} Dirac bispinor indices appearing for the introduced 
generalized Dirac fields and treated as undistinguishable physical degrees of 
freedom. However, in Refs. [11] it was wrongly presumed that two light sterile 
neutrinos mixed largely with two active neutrinos $\nu_e$ and $ \nu_\mu $, 
what now must be specifically corrected, of course.

The cosmological problems of light sterile neutrinos will not be discussed in 
this note.

\vspace{0.3cm} 

\ni {\bf 2. Overall neutrino mixing matrix}. In the present note, we will 
conjecture that two light sterile neutrinos, call them $\nu_s$ and $\nu_{s'}$, 
mix maximally with each other, but only moderately with three active neutrinos 
$\nu_e, \nu_\mu , \nu_\tau$. More precisely, we will assume the overall 
$5\times 5$ neutrino mixing matrix $U^{(5)}=\left(U^{(5)}_{\alpha i}\right)$ ~
($\alpha= e$, $\mu$, $\tau$, $s$, $s^\prime$ and $i=1, 2, 3,4,5$) in the form

\vspace{-0.2cm} 

%rownanie 3
\begin{equation}
U^{(5)} = U^{(5)}(12) U^{(5)}(14) = 
\left( \begin{array}{ccccc} 
c_{12}c_{14} & s_{12} & 0 & c_{12}s_{14} & 0 \\ 
-\frac{1}{\sqrt2}s_{12}c_{14}\;\;\, & \frac{1}{\sqrt2}c_{12} & 
\frac{1}{\sqrt2} & -\frac{1}{\sqrt2}s_{12}s_{14}\;\;\, & 0 \\ 
\frac{1}{\sqrt2}s_{12}c_{14} & -\frac{1}{\sqrt2}c_{12}\;\;\, & 
\frac{1}{\sqrt2} & \frac{1}{\sqrt2}s_{12}s_{14} & 0 \\
-\frac{1}{\sqrt2}s_{14}\;\;\, & 0 & 0 & \frac{1}{\sqrt2}c_{14} & 
\frac{1}{\sqrt2} \\
\frac{1}{\sqrt2}s_{14} & 0 & 0 & -\frac{1}{\sqrt2}c_{14}\;\;\, & 
\frac{1}{\sqrt2} \end{array} \right)  \,,
\end{equation}

\ni where 

\vspace{-0.2cm}

%rownanie 4
\begin{eqnarray}
U^{(5)}(12) & =\! & \left( \begin{array}{ccccc} 
c_{12} & s_{12} & 0 & 0 & 0 \\ 
-\frac{1}{\sqrt2}s_{12}\;\;\, & \frac{1}{\sqrt2}c_{12} & 
\frac{1}{\sqrt2}& 0 & 0 \\ 
\frac{1}{\sqrt2}s_{12} & -\frac{1}{\sqrt2}c_{12}\;\;\, & 
\frac{1}{\sqrt2} & 0 & 0 \\
0 & 0 & 0 & 1 & 0 \\
0 & 0 & 0 & 0 & 1 \end{array} \right) \,,  \nonumber \\ & & \nonumber \\
U^{(5)}(14) & = & \left( \begin{array}{ccccc} 
c_{14} & 0 & 0 & s_{14} & 0 \\ 
0 & 1 & 0 & 0 & 0 \\ 0 & 0 & 1 & 0 & 0 \\
-\frac{1}{\sqrt2}s_{14}\;\;\, & 0 & 0 & \frac{1}{\sqrt2}c_{14} & 
\frac{1}{\sqrt2} \\ 
\frac{1}{\sqrt2}s_{14} & 0 & 0 & -\frac{1}{\sqrt2}c_{14}\;\;\, & 
\frac{1}{\sqrt2} \\
\end{array} \right) \,.
\end{eqnarray}

\ni The first matrix factor in Eq. (3) arises from the bilarge form (2) of 
active-neutrino mixing matrix by its trivial $5\times 5$ extension, while 
the second is an analogue of the first, when $e$, $\mu$, 
$\tau\leftrightarrow e, s, s'$ and $1,2,3 \leftrightarrow 1,4,5$. Thus, the 
cosine $c_{14}$ and sine $s_{14}$  are analogues of cosine $c_{12}$ and sine 
$s_{12}$, though the angle $\theta_{14}$ is expected to be smaller than the 
large angle $\theta_{12} \sim 33^\circ $. Also $c_{45} = 1/\sqrt2 = s_{45}$ 
with the maximal angle $\theta_{45} = 45^\circ $ are analogues of 
$c_{23} = 1/\sqrt2 = s_{23}$ with the maximal  $\theta_{23} = 45^\circ $. 
Finally, an analogue of $s_{13} = 0$ is $s_{15} = 0$. Both kinds of conditions 
are necessary for the maximal mixing of $\nu_\mu$ with $\nu_\tau$ and $\nu_s$ 
with $\nu_{s^\prime}$.

The overall $5\times 5$ neutrino mixing matrix (3) leads to the following 
unitary transformation $\nu_i  = \sum_\alpha U^{(5)*}_{\alpha i}\,\nu_\alpha$ 
inverse to $\nu_\alpha  = \sum_i U^{(5)}_{\alpha i}\, \nu_i $:

%rownanie 5
\begin{eqnarray}
\nu_1  & = & c_{14}\left(c_{12}\nu_e  - s_{12} 
\frac{\nu_\mu - \nu_\tau}{\sqrt2}\right) - s_{14} 
\frac{\nu_s - \nu_{s'}}{\sqrt2} \;, \nonumber \\
\nu_2  & = & s_{12}\nu_e + c_{12} 
\frac{\nu_\mu - \nu_\tau}{\sqrt2} \;, \nonumber \\
\nu_3  & = & \frac{\nu_\mu + \nu_\tau}{\sqrt2} \;, \nonumber \\
\nu_4  & = & s_{14}\left(c_{12}\nu_e  - s_{12} 
\frac{\nu_\mu - \nu_\tau}{\sqrt2}\right) + c_{14}
\frac{\nu_s - \nu_{s'}}{\sqrt2} \;, \nonumber \\
\nu_5  & = & \frac{\nu_s + \nu_{s'}}{\sqrt2} \;. 
\end{eqnarray}

\ni This displays explicitly the maximal mixing of $\,\nu_\mu\,$ and 
$\,\nu_\tau\,$ as well as of $\,\nu_s\,$ and $\,\nu_{s'}$, because these 
neutrinos appear in Eq. (5) through the combinations 
$(\nu_\mu \mp \nu_\tau)/\sqrt2$ as well as $(\nu_s \mp \nu_{s'})/\sqrt2$, 
where $(\nu_\mu + \nu_\tau)/\sqrt2$ as well as $(\nu_s + \nu_{s'})/\sqrt2$ 
are decoupled from other flavor neutrinos (do not mix with them).

In the flavor representation, where the charged-lepton mass matrix is 
diagonal, the active-neutrino mixing matrix (2) is at the same time the 
diagonalizing matrix for the active-neutrino effective Majorana mass matrix. 
In this flavor representation, the overall $5\times 5$ effective neutrino 
mass matrix $M^{(5)} = \left( M^{(5)}_{\alpha \beta} \right)$  
$(\alpha, \beta = e, \mu, \tau, s, s')$ can be calculated from the formula

%rownanie 6
\begin{equation}
M^{(5)}_{\alpha \beta}  = \sum_i U^{(5)}_{\alpha i}\, m_i\, 
U^{(5)*}_{\beta i}\;,
\end{equation}

\ni where the matrix elements $U^{(5)}_{\alpha i}$ are given in Eq. (3). 
The form (6) is inverse to the diagonalization formula

%rownanie 7
\begin{equation}
\sum_{\alpha\, \beta} U^{(5)*}_{\alpha i}\, M^{(5)}_{\alpha \beta}\, 
U^{(5)}_{\beta j} = m_i \delta_{i j}\;.
\end{equation}

\vspace{0.3cm}

\ni {\bf 3. Overall neutrino oscillations}. We will use the 
$\nu_\alpha \rightarrow \nu_\beta$ neutrino oscillation probabilities 
(in the vacuum)

%rownanie 8
\begin{equation}
P(\nu_\alpha \rightarrow \nu_\beta) = 
\delta_{\beta \alpha} - 4\sum_{j>i} U^{(5)*}_{\beta j} U^{(5)}_{\alpha j}\, 
U^{(5)}_{\beta i} U^{(5)*}_{\alpha i} \sin^2 x_{j i}
\end{equation}

\ni $(\alpha, \beta = e, \mu, \tau, s, s'$ and $i,j = 1,2,3,4,5)$, where 

%rownanie 9
\begin{equation} 
x_{j i} \equiv 1.27 \frac{\Delta m^2_{j i} L}{E}\; , \; \Delta m^2_{j i} 
\equiv m^2_j - m^2_i
\end{equation}

\ni  ($\Delta m^2_{j i}$, $L$ and $E$ are measured in eV$^2$, km and GeV, 
respectively). Here, CP violation is neglected {\it i.e.}, 
$U^{(5)*}_{\alpha i} = U^{(5)}_{\alpha i}$ (or, more generally, the quartic 
products in Eq. (8) are real). For $U^{(5)}_{\alpha i}$ we will make use of 
the matrix elements of $U^{(5)}$ as given in Eq. (3).

The formula (8) applied respectively to the 
$\bar{\nu}_\mu \rightarrow \bar{\nu}_e $, $\nu_e \rightarrow \nu_e$ and 
$\nu_\mu \rightarrow \nu_\mu $ oscillations leads to the probabilities 
(in the vacuum)

%rownanie 10,11
\begin{eqnarray}
P(\bar{\nu}_\mu \rightarrow \bar{\nu}_e) & \simeq & 2c^2_{12}s^2_{12} 
c^2_{14} \sin^2 x_{21} + 2c^2_{12} s^2_{12}s^4_{14} \sin^2 x _{41} \,, \\
P(\nu_e \rightarrow \nu_e)& \simeq & 1 - 4c^2_{12}s^2_{12}s^2_{14} \sin^2 
x _{21} - 4c^2_{12}s^2_{14}\left(1  - c^2_{12}s^2_{14} \right) \sin^2 x _{41} 
\end{eqnarray}

\ni and

%rownanie 12
\begin{eqnarray}
P(\nu_\mu \rightarrow \nu_\mu) & \simeq & 1 - c^2_{12}s_{12}^2 c^2_{14} 
\sin^2 x _{21} - \left(1 - s^2_{12} s^2_{14}\right) \sin^2 x _{31} \nonumber \\
& &  - 2s^2_{12}s^2_{14}\left(1 - \frac{1}{2}s^2_{12} s^2_{14}\right)  
\sin^2 x _{41} \,, 
\end{eqnarray}

\ni when $x_{31}\simeq x_{32}$ and $x_{41} \simeq x_{42} \simeq x_{43}$ 
(notice that here, $x_{5i}$ and so, $m^2_5$ are absent). When, in addition, 
$x_{21} \ll |x_{31}| \ll x_{41}$ ({\it i.e.}, 
$m^2_1 < m^2_2 \ll m^2_3 \ll m^2_4$ or 
$m^2_3 \ll m^2_1 \simeq m^2_2 \ll m^2_4$) with  
$(x_{41})_{\rm LSND}= O(\pi /2)$, $(x_{21})_{\rm sol}= O(\pi /2)$, 
$(x_{31})_{\rm Chooz} \simeq (x_{31})_{\rm atm}= O(\pi /2)$ and 
$(x_{31})_{\rm atm}= O(\pi /2)$ for the LSND effect, for the solar 
$\nu_e$'s, for the Chooz reactor $\bar{\nu}_e$'s and for the atmospheric 
$\nu_\mu$'s, respectively, we get from Eqs. (10), (11) and (12) the 
probabilities (in the vacuum)

\vspace{-3mm}

%rownanie 13,14,15
\begin{eqnarray}
P(\bar{\nu}_\mu \rightarrow \bar{\nu}_e)_{\rm LSND} & \simeq & 
2c^2_{12} s^2_{12} s^4_{14} \sin^2 (x_{41})_{\rm LSND} \,, \\
P(\nu_e \rightarrow \nu_e)_{\rm sol}\;\;\;\; & \simeq & 
1 - 4c^2_{12}s^2_{12}c^2_{14} \sin^2 (x _{21})_{\rm sol} - 2c^2_{12}s^2_{14} 
\left(1 - c^2_{12}s^2_{14} \right) \,, \\ 
P(\bar{\nu}_e \rightarrow \bar{\nu}_e)_{\rm Chooz} & 
\simeq & 1 - 2c^2_{12}s^2_{14} \left(1 - c^2_{12} s^2_{14} \right)  
\end{eqnarray}

\ni and

\vspace{-2mm}

%rownanie 16
\begin{equation}
P(\nu_\mu \rightarrow \nu_\mu)_{\rm atm}  \simeq 
1 - \left(1 - s^2_{12}s^2_{14} \right) \sin^2 
(x_{31})_{\rm atm} -  s^2_{12}s^2_{14}  \left(1 - \frac{1}{2}s^2_{12}s^2_{14} 
\right)\,.
\end{equation}

\ni Of course, for solar $\nu_e$'s the MSW matter effect is significant, 
leading to the accepted LMA solar solution.

For the LSND effect of the order

\vspace{-3mm}

%rownanie 17
\begin{equation}
P(\bar{\nu}_\mu \rightarrow \bar{\nu}_e)_{\rm LSND} \sim  10^{-3} 
\sin^2 (x _{41})_{\rm LSND} 
\end{equation}

\ni and of the mass scale, say, $\Delta m^2_{41} \sim 1\;{\rm eV}^2$ we 
obtain the estimation

\vspace{-3mm}

%rownanie 18
\begin{equation} 
s^2_{14} \sim \left(\frac{10^{-3}}{2c^2_{12}s^2_{12}}\right)^{1/2} \sim 0.049
\end{equation}

\ni and so, $\theta_{14} \sim 13^\circ $, when $\theta_{12} \sim 33^\circ $. 
This implies the following estimates:

\vspace{-3mm}

%rownanie 19,20
\begin{eqnarray}
P(\nu_e \rightarrow \nu_e)_{\rm sol}\;\;\;\; & \sim & 1 - (0.83 - 0.041)\sin^2 
(x _{21})_{\rm sol} - 0.066 \,,  \\
P(\bar{\nu}_e \rightarrow \bar{\nu}_e)_{\rm Chooz} & \sim & 1 - 0.066 \,,
\end{eqnarray}

\ni and

\vspace{-3mm}

%rownanie 21
\begin{equation} 
P(\nu_\mu \rightarrow \nu_\mu)_{\rm atm} \sim 1 - (1 - 0.015) \sin^2 
(x _{31})_{\rm atm} - 0.014 \,. 
\end{equation} 

\ni Here, $ 4c^2_{12}s^2_{12} \sim 0.83$ ($c^2_{12} \sim 0.70$ and 
$s^2_{12} \sim 0.30$).

It can be noticed from Eq. (11) for $\nu_e \rightarrow \nu_e$ or 
$\bar{\nu}_e \rightarrow \bar{\nu}_e $ oscillations that the third 
mass-squared splitting $\Delta m^2_{\rm LSND} \equiv \Delta m^2_{41} 
\sim 1\;{\rm eV}^2$, characteristic for the reported LSND effect (Eq. (13)), 
may be manifested in principle also for $\nu_e \rightarrow \nu_e$ or 
$\bar{\nu}_e \rightarrow \bar{\nu}_e $ oscillations in any other experiment 
at an energy $E$ and a baseline $L$, where 
$(x_{41})_{\rm other} \simeq (x_{41})_{\rm LSND} = O(\pi /2)$ ({\it i.e.}, 
$(L/E)_{\rm other} \simeq (L/E)_{\rm LSND}$). In this case,

\vspace{-3mm}

%rownanie 22
\begin{eqnarray}
P(\nu_e \rightarrow \nu_e)_{\rm other} = 
P(\bar\nu_e \rightarrow \bar\nu_e)_{\rm other} & \!\simeq\! & 
1 - 4c^2_{12} s^2_{14} \left( 1 - c^2_{12}s^2_{14} \right) 
\sin^2 (x _{41})_{\rm other} \nonumber \\ & \!\sim\! & 1 - 0.13 
\sin^2 (x _{41})_{\rm other}\,,
\end{eqnarray}

\ni when  $\theta_{12} \sim 33^\circ $ and $s^2_{14} \sim 0.049$. Any such 
experiment might play for the LSND accelerator effect a somewhat similar role 
as that played by the Chooz experiment for the SuperKamiokande atmospheric 
experiment, where $(x_{31})_{\rm Chooz} \simeq (x_{31})_{\rm atm} = O(\pi /2)$ 
({\it i.e.}, $(L/E)_{\rm Chooz} \simeq (L/E)_{\rm atm}$). An analogical remark 
on $\Delta m^2_{41}$ may pertain also to Eq. (12) for 
$\nu_\mu \rightarrow \nu_\mu$ oscillations.

\vspace{0.3cm}

\ni {\bf 4. Conclusions}. When waiting for the test of LSND effect by the 
MiniBooNE experiment that may confirm or refute the LSND result, we presented 
in this note a 3+2 neutrino model, where two light sterile neutrinos mix 
maximally with each other and only moderately with three active neutrinos. 
The way of mixing is described by the $5\times 5$ mixing matrix (3).

Then, the LSND effect of the order (17) implies the estimates (19), (20) and 
(21) for the solar anomaly, Chooz negative result and atmospheric anomaly, 
respectively.

Note that in the conventional three-neutrino scheme (including, in general, 
$s_{13} \neq 0$) the result corresponding to Eq. (15) is

\vspace{-2mm}

%rownanie 23
\begin{equation}
P(\bar{\nu}_e \rightarrow \bar{\nu}_e)_{\rm Chooz} \simeq 
1 - 4 c^2_{13}s^2_{13} \sin^2(x_{31})_{\rm Chooz} \,,
\end{equation}

\ni when $x_{31} \simeq x_{32}$. From the negative result of Chooz experiment 
$4 c^2_{13}s^2_{13} < 0.12$ as $s^2_{13} < 0.03$ (here, $(x_{31})_{\rm Chooz} 
\simeq (x_{31})_{\rm atm} = O(\pi /2)$).

As is shown in Appendix A, the simplest 3+1 model with one light sterile 
neutrino leads to {\it the same} estimates (19), (20) and (21) as the 3+2 
model with two maximally mixing light sterile neutrinos, if the LSND effect 
is of the order (17). In fact, the oscillations (A7), (A8), (A9) and (A10) 
are {\it identical} with those given in Eqs. (13), (14), (15) and (16). 

Thus, the attractive 3+2 neutrino model with maximal mixing of two sterile 
neutrinos, presented in this note, is not {\it more} effective in explaining 
the hypothetic LSND result than the simplest 3+1 neutrino model. On the other 
hand, as is indicated in Appendix B, in the 3+2 model with maximal mixing of 
two sterile neutrinos where $s_{15} = 0$ (but $s_{14} \neq 0$), the deviations 
from  conventional oscillations of three active neutrinos (where $s_{14} = 0$ 
and $s_{15} = 0$) are {\it minimal} within a larger class of 3+2 models 
allowing for  $s_{15} \neq 0$ (in addition to $s_{14} \neq 0$).  

\vfill\eject

\vspace{0.2cm}

\def\theequation{A.\arabic{equation}}
\setcounter{equation}{0}

\centerline{\bf Appendix A:}

\centerline{\bf Comparing with the simplest 3+1 neutrino model}

\vspace{0.35cm}

Consider the simplest 3+1 neutrino model with one sterile neutrino, described 
by the overall $4\times 4$ mixing matrix bilarge in three active neutrinos:

%(A1)
\begin{eqnarray}
U^{(4)} = U^{(4)}(12) U^{(4)}(14) = 
\left( \begin{array}{cccc} 
c_{12}c_{14} & s_{12} & 0 & c_{12}s_{14} \\ 
-\frac{1}{\sqrt2}s_{12}c_{14}\;\;\, & \frac{1}{\sqrt2}c_{12} & 
\frac{1}{\sqrt2} & -\frac{1}{\sqrt2}s_{12}s_{14}\;\;\, \\ 
\frac{1}{\sqrt2}s_{12}c_{14} & -\frac{1}{\sqrt2}c_{12} & 
\frac{1}{\sqrt2}\;\;\, & \frac{1}{\sqrt2}s_{12}s_{14} \\
-s_{14}\;\;\, & 0 & 0 & c_{14} \\ \end{array} \right) \,, 
\end{eqnarray}

\ni where

%(A2)
\begin{eqnarray}
U^{(4)}(12) =\left( \begin{array}{rrrr} 
c_{12} & s_{12} & 0 & 0 \\ -\frac{1}{\sqrt2}s_{12} & \frac{1}{\sqrt2}c_{12} & 
\frac{1}{\sqrt2} & 0 \\ 
\frac{1}{\sqrt2}s_{12} & -\frac{1}{\sqrt2}c_{12} & \frac{1}{\sqrt2} & 0 \\ 
0 & 0 & 0 & 1 \\ \end{array} \right) \;,\;
U^{(4)}(14) = \left( \begin{array}{rrrr} c_{14} & 0 & 0 & s_{14} \\ 
0 & 1 & 0 & 0 \\  0 & 0 & 1 & 0 \\ 
-s_{14} & 0 & 0 & c_{14} \\ \end{array} \right) \,. 
\end{eqnarray}

\ni The cosine $c_{14}$ and sine $s_{14}$ correspond to an unknown mixing 
angle $\theta_{14}$ that has to be estimated from the reported LSND result.

The form (A1) of the mixing matrix $U^{(4)} = \left(U^{(4)}_{\alpha i}\right)$ 
($\alpha = e, \mu, \tau, s$ and $i = 1,2,3,4$) leads to the following unitary 
transformation $ \nu_i  = \sum_\alpha U^{(4)*}_{\alpha i}\, \nu_\alpha$ 
inverse to $\nu_\alpha  = \sum_i U^{(4)}_{\alpha i}\, \nu_i $:
%(A3)
\begin{eqnarray}
\nu_1  & = & c_{14}\left(c_{12}\nu_e  - s_{12} \frac{\nu_\mu - \nu_\tau}
{\sqrt2}\right) - s_{14}\nu_s \;, \nonumber\\
\nu_2  & = & s_{12}\nu_e + c_{12} \frac{\nu_\mu - \nu_\tau}{\sqrt2} \;, 
\nonumber \\
\nu_3  & = & \frac{\nu_\mu + \nu_\tau}{\sqrt2} \;, \nonumber \\
\nu_4  & = & s_{14}\left(c_{12}\nu_e  - s_{12} 
\frac{\nu_\mu - \nu_\tau}{\sqrt2}\right) + c_{14}\nu_s \;. 
\end{eqnarray}
This displays the maximal mixing of active neutrinos $\nu_\mu $ and 
$\nu_\tau $ which appear in the combinations $(\nu_\mu \mp \nu_\tau)/\sqrt2$, 
where $(\nu_\mu + \nu_\tau)/\sqrt2$ is decoupled from other flavor neutrinos, 
while the mixing of $\nu_e $ and $(\nu_\mu - \nu_\tau)/\sqrt2$ is large 
(though not maximal) having the mixing angle $\theta_{12} \sim 33^\circ $.

The mixing matrix (A1) implies the neutrino oscillation probabilities (in the 
vacuum)
%(A4,A5)
\begin{eqnarray}
P(\bar{\nu}_\mu \rightarrow \bar{\nu}_e) & \simeq & 
2 c^2_{12}s^2_{12}c^2_{14} \sin^2 x_{21} + 2 c^2_{12} s^2_{12} s^4_{14} 
\sin^2 x_{41} \,, \\
P(\nu_e \rightarrow \nu_e)\, & \simeq & 1 - 4 c^2_{12}s^2_{12}c^2_{14} 
\sin^2  x_{21} - 4 c^2_{12} s^2_{14}(1 - c^2_{12}s^2_{14}) \sin^2  x_{41} 
\end{eqnarray}
and
\begin{eqnarray}
P(\nu_\mu \rightarrow \nu_\mu) \simeq 1 \!-\! c^2_{12}s^2_{12}c^2_{14} 
\sin^2 x_{21} \!-\!(1 \!-\!s^2_{12} s^2_{14}) \sin^2 x_{31} \!-\!2s^2_{12} 
s^2_{14}(1-\frac{1}{2}s^2_{12}s^2_{14}) \sin^2 x_{41} \,,\nonumber\\ 
\end{eqnarray}

\ni when $x_{31} \simeq x_{32}$ and $x_{41} \simeq x_{42}  \simeq x_{43}$. 
Hence, when $x_{21} \ll |x_{31}| \ll x_{41}$ with $(x_{41})_{\rm LSND} 
= O(\pi /2)$, $(x_{21})_{\rm sol} = O(\pi /2)$, $(x_{31})_{\rm Chooz} = 
O(\pi /2)$ and $(x_{31})_{\rm atm} = O(\pi /2)$, respectively, for the LSND 
effect, for the solar $\nu_e$'s, for the Chooz reactor $\bar\nu_e$'s and for 
the atmospheric $\nu_\mu$'s, one obtains the probabilities (in the vacuum)
%(A7,A8,A9)
\begin{eqnarray}
P(\bar{\nu}_\mu \rightarrow \bar{\nu}_e)_{\rm LSND} & \simeq & 2 c^2_{12} 
s^2_{12} s^4_{14} \sin^2 (x_{41})_{\rm LSND} \,, \\
P(\nu_e \rightarrow \nu_e)_{\rm sol}\;\;\;\; & \simeq & 
1 - 4 c^2_{12}s^2_{12}c^2_{14} \sin^2(x_{21})_{\rm sol} 
- 2c^2_{12} s^2_{14}(1 - c^2_{12} s^2_{14}) \,, \\
P(\bar{\nu}_e \rightarrow \bar{\nu}_e)_{\rm Chooz} & 
\simeq & 1 - 2 c^2_{12}s^2_{14}(1 - c^2_{12} s^2_{14}) 
\end{eqnarray}
and
\begin{eqnarray}
P(\nu_\mu \rightarrow \nu_\mu)_{\rm atm} \simeq 
1 - (1- s^2_{12}s^2_{14}) \sin^2(x_{31})_{\rm atm} 
- s^2_{12} s^2_{14} \left(1 - \frac{1}{2}s^2_{12} s^2_{14}\right) \,. 
\end{eqnarray}

From Eq. (A7), for the LSND effect of the order 
$P(\bar{\nu}_\mu\rightarrow\bar{\nu}_e)_{\rm LSND}\sim10^{-3}
\sin^2 (x_{41})_{\rm LSND}$ and of the mass scale, say, 
$\Delta m^2_{41} \sim 1\;{\rm eV}^2$ one gets the estimation
\begin{eqnarray}
s^2_{14} \sim \left(\frac{10^{-3}}{2c^2_{12}s^2_{12}}\right)^{1/2} 
\sim 0.049 \,, 
\end{eqnarray}
when $\theta_{12} \sim 33^\circ $. This gives the following estimates:
\begin{eqnarray}
P(\nu_e \rightarrow \nu_e)_{\rm sol}\;\;\;\, & \sim & (1 - 0.83 - 0.041)
\sin^2 (x _{21})_{\rm sol} - 0.066\;,  \\
P(\bar{\nu}_e \rightarrow \bar{\nu}_e)_{\rm Chooz} & \sim & 1 - 0.066 
\end{eqnarray}
and
\begin{eqnarray}
P(\nu_\mu \rightarrow \nu_\mu)_{\rm atm}  \sim \;1 - (1- 0.015) 
\sin^2(x_{31})_{\rm atm}- 0.014 \;. 
\end{eqnarray}

We can see that the oscillations (A7)--(A10) and their estimates 
(A12)--(A14) in the case of $s^2_{14} \sim 0.049$, valid in the 
simplest 3+1 neutrino model, are identical with the oscillations 
(13)--(16) and their estimates (19)--(21) in the case of $s^2_{14}\sim0.049$, 
obtained in the 3+2 neutrino model with maximal mixing of two light sterile 
neutrinos. This identity is, of course, a consequence of the fact that 
$U^{(4)}_{\alpha i} = U^{(5)}_{\alpha i}$ and $U^{(5)}_{\alpha 5} = 0$ for 
$\alpha = e, \mu$ and $i = 1,2,3$, as it can be seen from Eqs. (A1) and (3). 
And this is true also for $\alpha = \tau$, the oscillations involving 
$\nu_\tau$ being identical in both cases.

\vfill\eject

\vspace{2mm}

\def\theequation{B.\arabic{equation}}
\setcounter{equation}{0}

\centerline{\bf Appendix B:}

\centerline{\bf Perturbing the maximal mixing of two sterile neutrinos}

\vspace{3mm}

The mixing matrix (3) in the 3+2 model with two maximally mixing light 
sterile neutrinos requires the trivial value $s_{15} = 0$ corresponding 
to $\theta_{15} = 0$. Now, introduce the nontrivial mixing angle 
$\theta_{15} \neq 0$, replacing the factor matrix $U^{(5)}(14)$ in Eq. (3) 
by the more general form
\begin{eqnarray}
U^{(5)}(14,15) = \left( \begin{array}{ccccc} 
c_{14}c_{15} & 0 & 0 & s_{14}c_{15} & s_{15} \\ 
0 & 1 & 0 & 0 & 0 \\
0 & 0 & 1 & 0 & 0 \\
\!\!\!-\frac{1}{\sqrt2}(s_{14} \!+\! c_{14}s_{15}) & 0 & 0 & 
\frac{1}{\sqrt2}(c_{14} \!-\! s_{14}s_{15})\;\, & \frac{1}{\sqrt2}c_{15} \\   
\frac{1}{\sqrt2}(s_{14} \!-\! c_{14}s_{15}) & 0 & 0 & 
\!\!\!-\frac{1}{\sqrt2}(c_{14} \!+\! s_{14}s_{15})\;\, & 
\frac{1}{\sqrt2}c_{15} \end{array} \right) 
\end{eqnarray}
which is the $5\times 5$ trivially extended canonical form of 
$3\times 3$ real unitary matrix for $\alpha = e, s, s'$ and $i = 1,4,5$ 
with $c_{45} = 1/\sqrt2 = s_{45}$. Then, the mixing matrix (3) transits 
into the new overall $5\times 5$ neutrino mixing matrix  
\begin{eqnarray}
\lefteqn{U^{(5)} \!=\! U^{(5)}(12) U^{(5)}(14,15)}\nonumber \\ & & \;= 
\left( \begin{array}{ccccc} 
c_{12}c_{14}c_{15} & s_{12} & 0 & c_{12}s_{14}c_{15} & c_{12}s_{15}\\  
\!\!\!
-\frac{1}{\sqrt2}s_{12}c_{14}c_{15} & \frac{1}{\sqrt2}c_{12} & 
\!\frac{1}{\sqrt2}\! & \!\!\!-\frac{1}{\sqrt2}s_{12}s_{14}c_{15} & 
\!\!\!-\frac{1}{\sqrt2}s_{12}s_{15} \\ 
\frac{1}{\sqrt2}s_{12}c_{14}c_{15} & \!\!\!-\frac{1}{\sqrt2}c_{12} & 
\!\frac{1}{\sqrt2}\! & \frac{1}{\sqrt2}s_{12}s_{14}c_{15} & 
\frac{1}{\sqrt2}s_{12}s_{15} \\
\!\!\!-\frac{1}{\sqrt2}(s_{14}+c_{14}s_{15}) & 0 & 0 & 
\frac{1}{\sqrt2}(c_{14}- s_{14}s_{15}) & \frac{1}{\sqrt2}c_{15} \\
\frac{1}{\sqrt2}(s_{14} - c_{14}s_{15}) & 0 & 0 & 
\!\!\!-\frac{1}{\sqrt2}(c_{14} + s_{14}s_{15}) & 
\frac{1}{\sqrt2}c_{15} \end{array} \right) \,.
\end{eqnarray}

Of course, for $s_{15} = 0$ the mixing matrix (B2) comes back to the form (3).

A consequence of the new mixing matrix is the following unitary 
transformation $ \nu_i  = \sum_\alpha U^{(5)*}_{\alpha i}\, \nu_\alpha $:
\begin{eqnarray}
\nu_1  & = & c_{14}\left[c_{15}\left(c_{12}\nu_e  - s_{12} 
\frac{\nu_\mu - \nu_\tau}{\sqrt2}\right) - s_{15} 
\frac{\nu_s + \nu_{s'}}{\sqrt2}\right] -  s_{14} 
\frac{\nu_s - \nu_{s'}}{\sqrt2}\;, \nonumber \\
\nu_2  & = & s_{12}\nu_e + c_{12} \frac{\nu_\mu - \nu_\tau}{\sqrt2} \;, 
\nonumber \\
\nu_3  & = & \frac{\nu_\mu + \nu_\tau}{\sqrt2} \;, \nonumber \\
\nu_4  & = & s_{14}\left[c_{15}\left(c_{12}\nu_e  - s_{12} 
\frac{\nu_\mu - \nu_\tau}{\sqrt2}\right) -  s_{15} 
\frac{\nu_s + \nu_{s'}}{\sqrt2} \right] + c_{14} 
\frac{\nu_s - \nu_{s'}}{\sqrt2} \;,  \nonumber\\
\nu_5  & = & s_{15}\left(c_{12}\nu_e - s_{12} 
\frac{\nu_\mu - \nu_\tau}{\sqrt2}\right) +  c_{15} 
\frac{\nu_s + \nu_{s'}}{\sqrt2} \;. 
\end{eqnarray}
We can see from Eqs. (B3) that here, the maximal mixing of 
$\nu_\mu$ and $\nu_\tau$ is maintained, while the maximal mixing 
of $\nu_s$ nd $\nu_{s'}$ is perturbed if $s_{15} \neq 0$ (beside 
$s_{14} \neq 0$).

The new mixing matrix (B2) leads to the neutrino oscillation probabilities 
(in the vacuum):
\begin{eqnarray}
P(\bar{\nu}_\mu \rightarrow \bar{\nu}_e)_{\rm LSND} & \simeq & 
2 c^2_{12} s^2_{12} s^2_{14}(s^2_{14} + s^2_{15}) 
\sin^2 (x_{41})_{\rm LSND} + c^2_{12}s^2_{12}s^4_{15} \,, \\
P(\nu_e \rightarrow \nu_e)_{\rm sol}\;\;\;\; & \simeq & 
1 - 4 c^2_{12}s^2_{12}(1 - s^2_{14} - s^2_{15}) 
\sin^2(x_{21})_{\rm sol} - 2c^2_{12}(s^2_{14} + s^2_{15})  \,, \\
P(\bar{\nu}_e \rightarrow \bar{\nu}_e)_{\rm Chooz} & 
\simeq & 1 - 2 c^2_{12}(s^2_{14} + s^2_{15}) 
\end{eqnarray}
and
\begin{eqnarray}
P(\nu_\mu \rightarrow \nu_\mu)_{\rm atm} \simeq 1 - \left[1- s^2_{12}(s^2_{14} 
+ s^2_{15})\right] \sin^2(x_{31})_{\rm atm} - s^2_{12}
(s^2_{14} + s^2_{15}) \,, 
\end{eqnarray}
when $x_{31}\simeq x_{32}$, $x_{41} \simeq x_{42} \simeq x_{43}$, 
$x_{51} \simeq x_{52} \simeq x_{53} \simeq x_{54}$ and 
$x_{21} \ll |x_{31}| \ll x_{41} \ll x_{51}$ with  
$(x_{41})_{\rm LSND}= O(\pi /2)$, $(x_{21})_{\rm sol}= O(\pi /2)$ and 
$(x_{31})_{\rm Chooz} \simeq (x_{31})_{\rm atm}= O(\pi /2)$. Here, the 
respective higher powers of $s^2_{14} \ll 1$ and $s^2_{15} \ll 1$ are 
{\it neglected}.

From Eq. (B4), for the LSND effect of the order 
$P(\bar{\nu}_\mu \rightarrow \bar{\nu}_e)_{\rm LSND} \sim  10^{-3} 
\sin^2 (x _{41})_{\rm LSND}$ one obtains the estimation 

%\vspace{-2mm}

%(B8)
$$
\left(s^4_{14} + s^2_{14}s^2_{15} + \frac{1}{2\sin^2(x_{41})_{\rm LSND}} 
s^4_{15} \right)^{\!\!1/2}\sim \left(\frac{10^{-3}}{2c^2_{12}s^2_{12}} 
\right)^{\!\!1/2} \sim 0.049 \,, \eqno({\rm B8})
$$%720

\ni when $\theta_{12} \sim 33^\circ $. Here, $\sin^2(x_{41})_{\rm LSND} 
\sim 1/2$ to 1. Thus, $s^2_{14} + s^2_{15} = (s^4_{14} + 2s^2_{14}s^2_{15} 
+ s^4_{15})^{1/2} > \;{\rm (lhs\; of\; Eq.\, (B8))}\;\sim 0.049$ if 
$s_{15} \neq 0$, while $s^2_{14} \sim 0.049$ if $s_{15} = 0$ (as is the case 
in the 3+2 model with maximal mixing of two sterile neutrinos). Hence, one 
can infer that the deviations from conventional oscillations of three active 
neutrinos (with $s_{14} = 0$ and $s_{15} = 0$), being proportional to 
$s^2_{14} + s^2_{15}$ in Eqs. (B5)--(B7), get larger magnitudes in the case 
of $s_{15} \neq 0$ (and $s_{14} \neq 0$) than in the case of $s_{15} = 0$ 
(but $s_{14} \neq 0$), where two sterile neutrinos mix maximally (leading to 
the estimates (19)--(21)). Thus, when $s_{15} = 0$, the 3+2 neutrino models 
defined in Eq. (B2) for various values of $s_{15}$ become {\it minimal} (in 
the sense of the discussed deviations). Such a minimal character of the 
deviations from conventional neutrino oscillations is connected, therefore, 
with the maximal mixing of two sterile neutrinos, realized if $s_{15} = 0$ 
(but $s_{14} \neq 0$).

The oscillation probabilities (B4)--(B7) are valid obviously in the option 
of {\it hierarchical} sterile neutrinos, where $m^2_4 \ll m^2_5$ implying 
$x_{41} \ll x_{51}$. Then, our conclusion of minimal character of deviations 
from conventional neutrino oscillations works for $s_{15} = 0$ (but 
$s_{14} \neq 0$). It turns out that in the opposite option of {\it degenerate} 
sterile neutrinos, where $m^2_4 \simeq m^2_5$ leading to 
$x_{41} \simeq x_{51}$ and $x_{54} \simeq 0$, the new mixing matrix (B2) 
provides exactly the oscillation probabilities of the form (13)--(16), where 
$s^2_{14}$ is replaced now by $s^2_{14} +s^2_{15}-s^2_{14}s^2_{15}$, equal 
approximately to $s^2_{14} + s^2_{15}$. In this degenerate option, the LSND 
effect of the order $P(\bar{\nu}_\mu \rightarrow \bar{\nu}_e)_{\rm LSND} 
\sim  10^{-3} \sin^2 (x _{41})_{\rm LSND}$ gives the estimation

%(B9)
$$
s^2_{14} + s^2_{15} - s^2_{14}s^2_{15} \sim 
\left(\frac{10^{-3}}{2c^2_{12}s^2_{12}} \right)^{\!\!1/2} \sim 0.049 \,, 
\eqno({\rm B9})
$$

\ni and the deviations from conventional oscillations of three active 
neutrinos are identical to those in the 3+2 model with two maximally 
mixing sterile neutrinos. They are equal to the previous minimal deviations 
appearing in the hierarchical option if $s_{15} = 0$ (but $s_{14} \neq 0$).

\vfill\eject

~~~~

\vspace{0.5cm}

{\centerline{\bf References}}

\vspace{0.5cm}

{\everypar={\hangindent=0.7truecm}
\parindent=0pt\frenchspacing

{\everypar={\hangindent=0.7truecm}
\parindent=0pt\frenchspacing

~[1]~Q.R. Ahmad {\it et al.} (SNO Collaboration), {\it Phys. Rev. Lett.} 
{\bf 87}, 071301 (2001); {\tt nucl--ex/0309004}.

\vspace{0.2cm}

~[2]~Y. Fukuda {\it et al.} (SuperKamiokande Collaboration), 
{\it Phys. Rev. Lett.} {\bf 81}, 1562 (1998); {\it Phys. Lett.} 
{\bf B 467}, 185 (1999).

\vspace{0.2cm}

~[3]~K. Eguchi {\it et al.} (KamLAND Collaboration), {\it Phys. Rev. Lett.} 
{\bf 90}, 021802 (2003).

\vspace{0.2cm}

~[4]~For a recent review {\it cf.} V. Barger, D. Marfatia and K. Whisnant, 
{\tt hep--ph/0308123}; M.~Maltoni, {\tt 
hep--ph/0401042}; and references therein.

\vspace{0.2cm}

~[5]~M. Apollonio {\it et al.} (Chooz Collaboration), {\it Eur. Phys. J.} 
{\bf C 27}, 331 (2003).

\vspace{0.2cm}

~[6]~C. Athanassopoulos {\it et al.} (LSND Collaboration), 
{\it Phys. Rev. Lett.} {\bf 77}, 3082 (1996); {\it Phys. Rev. } 
{\bf C 58}, 2489 (1998); A. Aguilar {\it et. al.}, {\it Phys. Rev.} 
{\bf D 64}, 112007 (2001).

\vspace{0.2cm}

~[7]~H. Murayama, and T. Yanagida, {\it Phys. Rev.} {\bf B 520}, 263 (2001); 
G. Borenboim, L.~Borissov, J. Lykken and A.Y. Smirnov, {\it JHEP} 
{\bf 0210}, 001 (2002); G. Borenboim, L. Borissov and J. Lykken, 
{\tt hep--ph/0212116}. However, the CPT violation scheme is disfavored by 
the atmospheric neutrino data, {\it cf.} A. Strumia, {\it Phys. Lett.} 
{\bf B 539}, 91 (2002).

\vspace{0.2cm}

~[8]~A.O. Bazarko {\it et al.}, {\tt hep--ex/9906003}.

\vspace{0.2cm}

~[9]~V. Barger, S. Pakvasa, T.J. Weiler and K. Whisnant, {\it Phys. Rev.} 
{\bf D 58}, 093016 (1998); {\it Phys. Lett.} {\bf B 437}, 107 (1998); and 
references therein; M. Maltoni, T. Schwetz, M.A.~Tortola and J.W. Valle, 
{\it Nucl. Phys.} {\bf B 643}, 321 (2002).

\vspace{0.2cm}

[10]~M. Sorel, J. Conrad and M. Shaevitz, {\tt hep--ph/0305255}; also 
K.S. Babu and G. Seidl, {\tt hep--ph/0312285}; K.L. McDonald, B.H.J. 
McKellar and A. Mastrano, {\tt hep--ph/0401241}.

\vspace{0.2cm}

[11]~W. Kr\'{o}likowski, {\it Acta Phys. Pol.} {\bf B 30}, 227 (1999), 
Appendix; in {\it Proc. of the 12th Max Born Symposium, Wroc{\l}aw, Poland, 
1998}, eds. A.~Borowiec {\it et al.}, Springer, 2000; {\it Acta Phys. Pol.} 
{\bf B 31}, 1913 (2000); also {\it cf. ibid.} {\bf B 33}, 2559 (2002). 

\vfill\eject

\end{document}